\documentstyle[aps,preprint]{revtex}

\begin{document}
\draft \tightenlines \preprint{gr-qc/9912114}

\title{Entropy Production and Thermodynamic Arrow of Time
in a Recollapsing Universe\footnote{Proceedings of the {\it 6th
Italian-Korean Meeting on Relativistic Astrophysics}, Pescara,
Italy, 1999}}

\author{Jung Kon Kim and Sang Pyo Kim}

\address{Department of Physics\\ Kunsan National University
\\Kunsan 573-701, KOREA}

\date{\today}

\maketitle
\begin{abstract}
We investigate the thermodynamic arrow of time in a
time-symmetrically recollapsing universe by calculating quantum
mechanically the entropy production of a massive scalar field. It
is found that even though the Hamiltonian has a time-reversal
symmetry with respect to the maximum expansion of the universe,
the entropy production is generic and the total entropy of the
scalar field increases monotonically. We conclude that the
thermodynamic arrow of time is a universal phenomenon even in the
expanding and subsequently recollapsing universe due to the
parametric interaction of matter field with gravity.
\end{abstract}
\pacs{42.50.D; 03.65.G;03.65.-w}
\date{\today}

\section{Introduction}

The thermodynamic arrow of time has always been a puzzling issue
in an expanding and subsequently recollapsing universe. When the
Universe is to recollapse time-symmetrically with respect to its
maximum expansion, can the entropy of matters always increase or
time-symmetrically decrease? Historically, Gold \cite{gold} gave
the first scientific explanation that the entropy of matter should
decrease time-symmetrically with respect to the maximum expansion
of the universe, but this argument was refuted by Penrose
\cite{penrose}. In the context of quantum cosmology, Hawking
\cite{hawking1} argued that the entropy of matter should be time
symmetric since the Hartle and Hawking's no-boundary wave function
\cite{hartle} should be CTP invariant. But Page \cite{page}
refuted this argument by pointing out the time-asymmetric wave
function obeying CTP theorem, in which the entropy can increase
monotonically throughout the expansion and subsequent recollapse
of the universe. Later Hawking {\it et al.} \cite{hawking2}
admitted the thermodynamic arrow of time by considering the
growing density perturbation even through the recollapsing
universe. Recently, we have argued that the parametric interaction
between the gravity and matter fields may break not only the
cosmological symmetry of time \cite{kim1} but also the
thermodynamic symmetry of time \cite{kim2}. Kiefer and Zeh
\cite{kiefer} also argued the thermodynamic arrow of time in the
recollapsing universe. (See Ref. \cite{halliwell} for
comprehensive review and references.)

In this paper we revisit the issue of thermodynamic arrow of time
by studying a massive scalar field in a time-symmetrically
recollapsing universe. Matter field in such a time-dependent
background spacetime should be described by a nonequilibrium
quantum field. We apply the recently introduced unified approach
\cite{kim3} based on the Liouville-von Neumann (LvN) equation to
the nonequilibrium massive scalar field and find the
time-dependent Fock (Hilbert) spaces throughout the expansion and
subsequent recollapse of the universe. The particle production is
calculated explicitly and the information-theoretic entropy is
computed for the nonequilibrium quantum process. From this we
conclude that the thermodynamic arrow of time is a universal
phenomenon due to the parametric interaction of the matter field
with the gravity even in the recollapsing universe.

The organization of this paper is as follows. In Sec. II we review
the nonequilibrium quantum evolution of the massive scalar field
in the recollapsing Friedmann-Robertson-Walker (FRW) universe. In
Sec. III we study an exactly solvable model for the massive scalar
field in the time-symmetrically recollapsing universe. Finally, in
Sec. IV we study the thermodynamic arrow of time in the
recollapsing FRW universe.

\section{Massive Scalar Field in the FRW Universe}

In order to study the thermodynamic arrow of time in the
time-symmetrically recollapsing universe, we consider the simple
model of a closed FRW universe minimally coupled to a massive
scalar field. The massive scalar field is described by the action
\begin{equation}
I = - \frac{1}{2} \int d^4x \sqrt{-g} \Bigl[g^{\mu \nu}
\partial_{\mu} \Phi \partial_{\nu} \Phi + m^2 \Phi^2 \Bigr].
\label{action}
\end{equation}
The FRW universe has the metric
\begin{equation}
ds^2 = - dt^2 + R^2 (t) d\Omega_3^2,
\end{equation}
where $d\Omega^2_3$ is the metric on the unit three-sphere.
According to Ref. \cite{lifschitz} the scalar field can be
decomposed as
\begin{equation}
\Phi (t, {\bf x}) = \sum_{n, l, m} \phi^n_{l,m} (t) Q^n_{l,m}
({\bf x}),
\end{equation}
where $Q^{n}_{l,m}$ denote the eigenfunctions of the Laplace
operator on the three-sphere and are explicitly given by $
Q^n_{l,m} = \Pi^n_l Y_{l,m}$, where $Y_{l,m}$ are the spherical
harmonics on the two-sphere and $\Pi^n_l$ are the Fock harmonics.
Then from the action (\ref{action}) we obtain the mode-decomposed
Hamiltonian for the scalar field
\begin{equation}
H (t) = \sum_{\bf k} H_{\bf k} (t) =: \sum_{\bf k} \frac{\pi_{\bf
k}^2}{2R^3(t)} + \frac{R^3(t)}{2} \Bigl(m^2 + \frac{n^2
-1}{R^2(t)} \Bigr) \phi^2_{\bf k}, \label{hamiltonian}
\end{equation}
where ${\bf k}$ denote collectively $(n, l, m)$ and $\pi_{\bf k}$
are momenta conjugate to $\phi_{\bf k}$. Thus, the Hamiltonian is
a collection of infinitely many time-dependent harmonic
oscillators.

Each time-dependent harmonic oscillator in Eq. (\ref{hamiltonian})
describes a nonequilibrium quantum process through the parametric
interaction between the matter and gravity. Such time-dependent
system leads necessarily to the particle production in quantum
mechanics and its nonequilibrium process implies the entropy
production in quantum statistical mechanics. To treat the
nonequilibrium process appropriately, we shall adopt the recently
introduced unified approach \cite{kim3} based on the LvN equation.
This LvN approach provides us explicitly not only with the exact
quantum states but also with the density operator. In this LvN
approach the fundamental laws for the nonequilibrium quantum
system are the time-dependent Schr\"{o}dinger equation (in unit
$\hbar$ = 1)
\begin{equation}
i \frac{\partial \Psi (t)}{\partial t} = \hat{H} (t) \Psi (t),
\label{ev eq}
\end{equation}
and the LvN equation for the density operator
\begin{equation}
i \frac{\partial \hat{\rho} (t)}{\partial t} + [ \hat{\rho} (t),
\hat{H} (t)] = 0. \label{d-ln eq}
\end{equation}
It was Lewis and Riesenfeld \cite{lewis} who observed that the
eigenstate of any operator ${\cal O}$ satisfying the LvN equation
(\ref{d-ln eq}) is an exact quantum state of the time-dependent
quantum system up to a trivial time-dependent phase factor. (For
development and various applications, see Ref. \cite{kim4}.)

The stratagem of the LvN approach is to find the pair of
time-dependent operators
\begin{equation}
\hat{A}_{\bf k} (t) = i \Bigl[ u_{\bf k}^* (t) \hat{\pi}_{\bf k} -
R^3 (t) \dot{u}_{\bf k}^* (t) \hat{\phi}_{\bf k} \Bigr],~~
\hat{A}_{\bf k}^{\dagger} (t) = {\rm h. c.} \Bigl(\hat{A}_{\bf k}
(t)\Bigr),
\end{equation}
that satisfy the LvN equation
\begin{equation}
i \frac{\partial \hat{A}_{\bf k} (t)}{\partial t}  + [
\hat{A}_{\bf k} (t), \hat{H}_{\bf k} (t) ] = 0. \label{ln eq}
\end{equation}
It readily follows that Eq. (\ref{ln eq}) is satisfied only when
$u_{\bf k}$ is a complex solution to the corresponding classical
equation of motion
\begin{equation}
\ddot{u}_{\bf k} (t) + 3 \frac{\dot{R}(t)}{R(t)} \dot{u}_{\bf k}
(t) + \Bigl(m^2 + \frac{n^2 -1}{R^2(t)} \Bigr) u_{\bf k} (t) = 0.
\label{cl eq1}
\end{equation}
$\hat{A}_{\bf k}(t)$ and $\hat{A}^{\dagger}_{\bf k'}(t)$ can be
made the annihilation and creation operators with the standard
commutation relation at each time $t$
\begin{equation}
[\hat{A}_{\bf k} (t), \hat{A}^{\dagger}_{\bf k'} (t) ] =
\delta_{\bf k, k'}
\end{equation}
by imposing the Wronskian condition
\begin{equation}
R^3 (t) \Bigl[u_{\bf k} (t) \dot{u}_{\bf k}^* (t) -  u^*_{\bf k}
(t) \dot{u}_{\bf k} (t) \Bigr] = i. \label{wronskian}
\end{equation}
Then, neglecting trivial time-dependent phase factors, the number
states
\begin{equation}
\vert n_{\bf k}, t \rangle = \frac{1}{\sqrt{n_{\bf k}!}} \Bigl(
\hat{A}_{\bf k} (t) \Bigr)^{n_{\bf k}} \vert 0_{\bf k}, t
\rangle,\label{num st}
\end{equation}
together with the vacuum state
\begin{equation}
\hat{A}_{\bf k} (t) \vert 0_{\bf k}, t \rangle = 0,
\label{gaussian}
\end{equation}
constitute the time-dependent Fock (Hilbert) space. In fact, there
are one-parameter Gaussian states (\ref{gaussian}) for each
oscillator and the vacuum state is the Gaussian state having the
minimum uncertainty \cite{kim4}.

Note that the Hamiltonian system (\ref{hamiltonian}) may not be
solved exactly since the classical equations of motion (\ref{cl
eq1}) can not be solved in closed form for the scale factor
\begin{equation}
R (t) = \sqrt{\frac{3m_P^2}{8 \pi \rho}} {\rm csch} \Bigl(
\sqrt{\frac{8 \pi \rho}{3m_P^2}}t \Bigr), \label{exp}
\end{equation}
where $m_P = \frac{1}{\sqrt{G}}$ is the Planck mass and $\rho$ is
the density of dust particles in the the closed FRW universe. So
it would be interesting to study first a model that can be solved
exactly.

\section{An Exactly Solvable Model}

We now consider a model Hamiltonian that shares many physically
interesting features with Eq. (\ref{hamiltonian}). The frequencies
of oscillators in Eq. (\ref{hamiltonian}) increase indefinitely as
the universe (re)collapses $(R \rightarrow 0)$, and approach to
constant values at the maximum expansion. To study the particle
production in a curved spacetime, the following Hamiltonian has
been often used \cite{birrel}
\begin{equation}
\hat{H} (t) = \sum_{\bf k} \hat{H}_{\bf k} (t)  =: \sum_{\bf k}
\frac{\hat{p}_{\bf k}^2}{2m_0} + \frac{m_0}{2} \Bigl( \omega_0^2 +
\frac{\omega_{\bf k}^2}{4}t^2\Bigr) \hat{q}_{\bf
k}^2,\label{model}
\end{equation}
whose the classical equations of motion can be solved analytically
\begin{equation}
\ddot{v}_{\bf k} (t) +  \Bigl( \omega_0^2 + \frac{\omega_{\bf
k}^2}{4}t^2\Bigr) v_{\bf k} (t) = 0. \label{cl eq2}
\end{equation}
In terms of the complex solution for each oscillator one finds the
annihilation and creation operators
\begin{equation}
\hat{B}_{\bf k} (t) = i \Bigl[ v_{\bf k}^* (t) \hat{p}_{\bf k} -
m_0 \dot{v}_{\bf k}^* (t) \hat{q}_{\bf k} \Bigr],  ~~\hat{B}_{\bf
k}^{\dagger} (t) = {\rm h.c.} \Bigl(\hat{B}_{\bf k} (t) \Bigr).
\label{op2}
\end{equation}
Further, one should impose the Wronskian condition
\begin{equation}
m_0 \Bigl[v_{\bf k} (t) \dot{v}_{\bf k}^* (t) -  v^*_{\bf k} (t)
\dot{u}_{\bf k} (t) \Bigr] = i, \label{wron2}
\end{equation}
to guarantee the standard commutation relation $[\hat{B}_{\bf k}
(t), \hat{B}^{\dagger}_{\bf k'} (t) ] = \delta_{\bf k, k'}$ for
all times.

The complex solution to Eq. (\ref{cl eq2}) satisfying the
Wronskian (\ref{wron2}) is given by
\begin{equation}
v_{{\bf k},in} (t) = \frac{1}{(4 m_0^2 \omega_{\bf k})^{1/4}}
\Bigl[i \sqrt{\kappa} W (a_{\bf k}, \tau) +
\frac{1}{\sqrt{\kappa}} W \Bigl(a_{\bf k}, - \tau) \Bigr]
\label{sol2}
\end{equation}
where $W (a, \pm \tau)$ are the parabolic cylinder functions
\cite{abramowitz}, and
\begin{equation}
a_{\bf k} = - \frac{\omega_0^2}{\omega_{\bf k}}, ~ \tau =
\sqrt{\omega_{\bf k}} t, ~ \kappa = \sqrt{1 + e^{2 \pi a_{\bf k}}}
- e^{\pi a_{\bf k}}.
\end{equation}
In the past infinity $(t \rightarrow - \infty)$ the solution
(\ref{sol2}) has the asymptotic form \cite{abramowitz}
\begin{equation}
v_{{\bf k}, in} (t) = \frac{1}{(m_0^2 \omega_{\bf k}
\tau^2)^{1/4}} \exp\Bigl[i \Bigl(\frac{\tau^2}{4} - a_{\bf k} \ln
(-\tau) + \frac{\pi}{4} + \frac{\varphi_2}{2} \Bigr)
\Bigr],\label{asymptotic}
\end{equation}
where $\varphi_2 = {\rm arg} \Gamma \Bigl(\frac{1}{2} + i a_{\bf
k} \Bigr)$. Note that the asymptotic solution (\ref{asymptotic})
will be the same as the adiabatic (WKB) solution of Sec. IV. Also
another useful form of the solution (\ref{sol2}) is given by
\begin{equation}
v_{{\bf k},in} (t) = \frac{1}{(4 m_0^2 \omega_{\bf k} )^{1/4}}
\Biggl[\frac{i}{2} \Bigl(k - \frac{1}{k} \Bigr) E(a_{\bf k}, \tau)
+ \frac{i}{2} \Bigl(k + \frac{1}{k} \Bigr) E^*(a_{\bf k}, \tau)
\Biggr].
\end{equation}
In the future infinity $(t \rightarrow \infty)$, $u_{{\bf k}, out}
$ has the asymptotic form
\begin{eqnarray}
v_{{\bf k}, out} &=& \frac{1}{(4 m_0^2 \omega_{\bf k} )^{1/4}} E^*
(a_{\bf k}, \tau) \nonumber\\ &=&  \frac{1}{(4 m_0^2 \omega_{\bf
k} \tau^2 )^{1/4}} \exp\Bigl[-i \Bigl(\frac{\tau^2}{4} - a_{\bf k}
\ln (\tau) + \frac{\pi}{4} + \frac{\varphi_2}{2} \Bigr) \Bigr].
\label{asymptotic2}
\end{eqnarray}

From Eqs. (\ref{asymptotic}) and (\ref{asymptotic2}) we are able
to find the transformation of solutions between two asymptotic
regimes
\begin{equation}
v_{{\bf k},in} (t) =  \frac{i}{2} \Bigl(\kappa + \frac{1}{\kappa}
\Bigr) v_{{\bf k}, out} (t) + \frac{i}{2} \Bigl(\kappa -
\frac{1}{\kappa} \Bigr) v^*_{{\bf k}, out} (t),
\end{equation}
and the Bogoliubov transformation
\begin{equation}
\hat{B}_{{\bf k},out} (t) =  \mu_{\bf k} \hat{B}_{{\bf k}, in} (t)
+ \nu_{\bf k} \hat{B}^{\dagger}_{{\bf k}, in} (t), ~~{\rm h.c.}
\label{bog tran}
\end{equation}
where
\begin{equation}
\mu_{\bf k} =  \frac{i}{2} \Bigl(\kappa + \frac{1}{\kappa} \Bigr),
~ \nu_{\bf k} = - \frac{i}{2} \Bigl(\kappa - \frac{1}{\kappa}
\Bigr).
\end{equation}
Also the Bogoliubov transformation can be rewritten as
\begin{equation}
\hat{B}_{{\bf k},out} (t) = \hat{S}_{\bf k} (z_{\bf k})
\hat{B}_{{\bf k},in} (t) \hat{S}^{\dagger}_{\bf k} (z_{\bf k}),
~{\rm h.c.} \label{unit tran}
\end{equation}
in terms of the squeeze operator \cite{yuen}
\begin{equation}
\hat{S}_{\bf k} (z_{\bf k}) = \exp \Bigl[i \frac{\pi}{2}
\hat{B}^{\dagger}_{{\bf k}, in} \hat{B}_{{\bf k}, in} \Bigr] \exp
\Bigl[\frac{r_{\bf k}}{2} \bigl( \hat{B}^2_{{\bf k}, in} -
\hat{B}^{\dagger 2}_{{\bf k},in} \bigr) \Bigr] ,
\end{equation}
in which
\begin{equation}
\mu_{\bf k} = \cosh r_{\bf k} e^{i \frac{\pi}{2}},~\nu_{\bf k} =
\sinh r_{\bf k} e^{i \frac{3\pi}{2}}
\end{equation}

The vacuum states (\ref{gaussian}) for each oscillator depend on
time explicitly and those at two different times are related by
the Bogoliubov transformation
\begin{equation}
\hat{B}_{\bf k} (t) =  \mu_{\bf k} (t, t_0) \hat{B}_{\bf k} (t_0)
+ \nu_{\bf k} (t, t_0) \hat{B}^{\dagger}_{\bf k} (t_0), ~~{\rm
h.c.}
\end{equation}
where
\begin{eqnarray}
\mu_{\bf k} (t, t_0) &=& i m_0 \bigl[ v^*_{\bf k} (t) \dot{v}_{\bf
k} (t_0) - \dot{v}^*_{\bf k} (t) v_{\bf k} (t_0) \bigr],
\nonumber\\\nu_{\bf k} (t, t_0) &=& i m_0 \bigl[ v^*_{\bf k} (t)
\dot{v}^*_{\bf k} (t_0) - \dot{v}^*_{\bf k} (t) v^*_{\bf k} (t_0)
\bigr].
\end{eqnarray}
The quantum state of each oscillator evolves unitarily by the
unitary transformation (\ref{unit tran}), and the vacuum states at
two different times are not orthogonal to each other
\begin{equation}
\langle 0_{\bf k}, t \vert 0_{\bf k}, t_0 \rangle = \Biggl(
\frac{e^{i\vartheta_{\bf k} (t, t_0)}}{\vert \mu_{\bf k} (t,
t_0)\vert} \Biggr)^{1/2},
\end{equation}
where $\vartheta_{\bf k} = {\rm arg} \bigl( v_{\bf k} (t_0) v_{\bf
k} (t) \bigr)$. However, since $\mu_{\bf k} \geq 1$ and can be
unity at most for a finite number of modes, the vacuum states for
all modes of the scalar field
\begin{equation}
\vert 0, t \rangle = \prod_{\bf k} \vert 0_{\bf k}, t \rangle,
\end{equation}
are orthogonal to each other at two different times
\begin{equation}
\langle 0, t \vert 0, t_0 \rangle = \prod_{\bf k} \Biggl(
\frac{e^{i \vartheta_{\bf k}(t, t_0)}}{\vert \mu_{\bf k} (t, t_0)
\vert} \Biggr)^{1/2} = 0.
\end{equation}
Therefore, the time-dependent Fock spaces constructed from the
vacuum states form unitary inequivalent representations. The
physical implication of the unitary inequivalence of
representations is that particles are produced in the infinite
future $(t \rightarrow \infty)$ from the initial vacuum state of
the infinite past $(t \rightarrow - \infty)$ by the amount
\begin{equation}
{}_{in}\langle 0 \vert \sum_{\bf k} \hat{B}^{\dagger}_{{\bf k},
out} \hat{B}_{{\bf k}, out} \vert 0 \rangle_{in} = \sum_{\bf k}
\nu^2_{\bf k}.
\end{equation}
Similarly, the particle production from the number states is found
to be
\begin{equation}
\Bigl[\prod_{\bf k} {}_{in}\langle n_{\bf k}  \vert \Bigr]
\sum_{\bf k} \hat{B}^{\dagger}_{{\bf k}, out}  \hat{B}_{{\bf k},
out}  \Bigl[\prod_{\bf k} \vert n_{\bf k}  \rangle_{in} \Bigr] =
\sum_{\bf n_{\bf k}} \nu^2_{\bf k} (2n_{\bf k} + 1).
\end{equation}
The particle production in the expanding universe was first
explained by Parker \cite{parker}.

We can make use of the operators (\ref{op2}) satisfying the LvN
equation to construct a density operator
\begin{eqnarray}
\hat{\rho}_{\bf k} (t) &=& \frac{1}{Z_{\bf k} (t)} e^{ - \beta
\Omega_{\bf k} \Bigl(\hat{B}^{\dagger}_{\bf k} (t) \hat{B}_{\bf k}
(t) + \frac{1}{2} \Bigr)}, \nonumber\\ Z_{\bf k} (t) &=& {\rm Tr}
e^{ - \beta \Omega_{\bf k} \Bigl( \hat{B}^{\dagger}_{\bf k} (t)
\hat{B}_{\bf k} (t) + \frac{1}{2} \Bigr)}.
\end{eqnarray}
Here $\beta$ is a free parameter for nonequilibrium quantum
processes, which becomes $\beta = \frac{1}{k_B T}$ for an
equilibrium system, where $k_B$ and $T$ are the Boltzmann constant
and temperature, respectively. Using the information-theoretic
entropy at each time defined by \cite{tolman}
\begin{equation}
S_{{\bf k}, ent.} (t) = - k_B {\rm Tr} \Bigl[ \hat{\rho}_{\bf k}
(t) \ln \hat{\rho}_{\bf k} (t) \Bigr], \label{ent}
\end{equation}
the entropy production during the evolution from the infinite past
to the infinite future is given by
\begin{equation}
\Delta S_{{\bf k}, ent.} = - k_B  \Biggl\{{\rm Tr} \Bigl[
\hat{\rho}_{{\bf k}, out}
 \ln \hat{\rho}_{{\bf k}, out} \Bigr] - {\rm Tr} \Bigl[
\hat{\rho}_{{\bf k}, in}  \ln \hat{\rho}_{{\bf k}, in} \Bigr]
\Biggr\}.\label{ent ch}
\end{equation}
It should be pointed out that the entropy cannot be produced for a
system of finite oscillators due to the unitary transformation
(\ref{unit tran}).

However, the Fock space of the scalar field in the infinite future
is unitarily inequivalent to that of the infinite past. In other
words, Eq. (\ref{unit tran}) should not be regarded as a unitary
transformation but rather as a unitarily inequivalent
transformation of the scalar field. Therefore, the entropy
production for the scalar field should be evaluated with respect
to the same Fock space, for instance, that of the infinite past
\begin{eqnarray}
\Delta S_{{\bf k}, ent.} &=& - k_B \sum_{n_{\bf k}} \Biggl\{
{}_{in}\langle n_{\bf k} \vert \hat{\rho}_{out}  \ln
\hat{\rho}_{out} \vert n_{\bf k} \rangle_{in} -  {}_{in}\langle
n_{\bf k} \vert \hat{\rho}_{in}  \ln \hat{\rho}_{in} \vert n_{\bf
k} \rangle_{in} \Biggr\} \nonumber\\ &=& - k_B \sum_{n_{\bf k},
m_{\bf k}} \Biggl\{{}_{in}\langle n_{\bf k} \vert \hat{\rho}_{in}
\ln \hat{\rho}_{in} \vert n_{\bf k} \rangle_{in} \Bigl( \vert
{}_{in}\langle m_{\bf k} \vert \hat{S} \vert n_{\bf k}
\rangle_{in} \vert^2 - \delta_{m_{\bf k}, n_{\bf k}} \Bigr)
\Biggr\}. \label{trace}
\end{eqnarray}
Here we made use of the transformation (\ref{unit tran}) in the
second line. For a small squeeze parameter $r_{\bf k}$ we obtain
approximately $\Delta S_{{\bf k}, ent.} \simeq 2 r_{\bf k} S_{{\bf
k}, in}$, which is different from $\Delta S_{{\bf k}, ent.} \simeq
2 r_{\bf k}$ for the Gaussian states in Refs. \cite{hu,grishchuk}.
The total amount of the information-theoretical entropy is given
by
\begin{equation}
\Delta S_{ent.} \simeq  \sum_{\bf k} 2 \ln \Bigl(\sqrt{1 + e^{- 2
\pi \frac{\omega_0^2}{\omega_{\bf k}}}} - e^{- \pi
\frac{\omega_0^2}{\omega_{\bf k}}} \Bigr) S_{{\bf k}, in}.
\end{equation}
It is remarkable that the entropy increases monotonically during
the symmetric evolution from the infinite past to the future and
the amount of the entropy production is proportional to the
entropy present in the past.

\section{Thermodynamic Arrow of Time in the Recollapsing
Universe}

We now return to the scalar field in the FRW universe and its
Hamiltonian (\ref{hamiltonian}). Ofttimes the instantaneous
annihilation and creation operators are used that diagonalize the
Hamiltonian at every moment:
\begin{equation}
\hat{A}_{{\bf k}, dig.} (t) = e^{i \int \Omega_{\bf k} (t)}
\Bigl[\sqrt{\frac{R^3 (t) \omega_{\bf k} (t)}{2}} \hat{q}_{\bf k}
+ i \sqrt{\frac{1}{R^3 (t) \omega_{\bf k}(t)}} \hat{p}_{\bf k}
\Bigr], ~~ \hat{A}^{\dagger}_{{\bf k}, dig.} (t) = {\rm h.c.}
\Bigl(\hat{A}_{{\bf k}, dig.} (t) \Bigr). \label{inst op}
\end{equation}
However, the Fock spaces of these number states lead to the
infinite production of particle and entropy \cite{birrel}. On the
other hand, the LvN approach in Secs. II and III will yield the
correct and physically meaningful result, provided that the
analytical solution of Eq. (\ref{cl eq1}) is used for each mode.
Since it is difficult to solve analytically Eq. (\ref{cl eq1}) for
the scale factor (\ref{exp}), we rely on the adiabatic (WKB)
solution. This is done by interpreting Eq. (\ref{cl eq1}) as a
one-dimensional Schr\"{o}dinger equation and, for a slowly varying
frequency $\omega_{\bf k} (t)$, by using the adiabatic solution of
the form
\begin{equation}
u_{{\bf k}, ad} (t) = \frac{1}{\sqrt{2 R^3 (t) \Omega_{\bf k}
(t)}} \exp \Bigl[- i \int \Omega_{\bf k} (t) \Bigr].\label{ad sol}
\end{equation}
The adiabatic annihilation and creation operators for each
oscillator are related with the instantaneous ones via the
Bogoliubov transformation
\begin{equation}
\hat{A}_{{\bf k}, ad} (t) = \Bigl(1 - \frac{i}{4 m_0}
\frac{\dot{\Omega}_{\bf k}}{\Omega^2_{\bf k}} \Bigr)\hat{A}_{{\bf
k}, dig.} (t) - \frac{i}{4 m_0} \frac{\dot{\Omega}_{\bf
k}}{\Omega^2_{\bf k}}e^{2 i \int \Omega_{\bf k} (t)}
\hat{A}^{\dagger}_{{\bf k}, dig.} (t),~~ {\rm h.c.}.
\end{equation}

As for the model in Sec. III, the adiabatic solution of the
infinite past can be written in terms of that of the infinite
future as
\begin{equation}
u_{{\bf k}, ad. in} = \mu_{\bf k} u_{{\bf k}, ad. out} - \nu_{\bf
k} u^*_{{\bf k}, ad. out}.
\end{equation}
The explicit forms of $\mu_{\bf k}$ and $\nu_{\bf k}$ depend on $R
(t) $, $m$, and ${\bf k}$. Similarly, the adiabatic solution at
$t_p$ in the expanding stage can be expressed in terms of that at
$t_f$ in the time-symmetric recollapsing stage as
\begin{equation}
u_{{\bf k}, ad.} (t_p) = \mu_{\bf k} (t_f, t_p) u_{{\bf k}, ad.}
(t_f) - \nu_{\bf k} (t_f, t_p) u^*_{{\bf k}, ad.} (t_f). \label{sc
tran}
\end{equation}
In general, one has $\mu_{\bf k} (t_f, t_p) \neq 0$ even for
symmetric times $t_p$ and $t_f$ with respect to the maximum
expansion of the universe. One can also understand this fact from
the scattering of a particle by a symmetric potential in quantum
mechanics. From the study of the model in Sec. III we may find
some characteristic features of Eq. (\ref{hamiltonian}). It is
rather straightforward to obtain the entropy production for each
mode of the scalar field
\begin{equation}
\Delta S_{{\bf k}, ent.} (t_f, t_p) \simeq 2 r_{\bf k} (t_f, t_p)
S_{{\bf k}} (t_p),
\end{equation}
where
\begin{equation}
\cosh {r_{\bf k}} (t_f, t_p) = \vert \mu_{\bf k} (t_f, t_p) \vert.
\end{equation}
The total entropy production of the scalar field is then given by
\begin{equation}
\Delta S_{ent.} (t_f, t_p) \simeq  \sum_{\bf k} 2 r_{\bf k} (t_f,
t_p) S_{{\bf k}} (t_p).
\end{equation}
Thus we have shown that the entropy of the matter (scalar) field
increases even during the time-symmetric recollapse of the
universe. This thermodynamic arrow of time is due to the
parametric interaction between the matter field and gravity. The
particle and entropy production seems to be a generic feature and
thereby, the global thermodynamic arrow of time is a consequence
of the evolution of the universe.

\section{Conclusion}

In this paper we have studied the particle and entropy production
of a massive scalar field in a time-symmetrically recollapsing
universe. In the expanding and subsequently recollapsing universe
the matter field is described by a nonequilibrium quantum field
due to the parametric interaction of the matter field with the
gravity. To treat properly we have applied the unified approach
based on the Liouville-von Neumann equation to the nonequilibrium
massive scalar field and found the time-dependent Fock spaces
throughout the evolution of the universe. It is the parametric
interaction that leads to the entropy production during the
evolution of the universe. The study of an exactly solvable model
which has many similar features with the scalar field revealed the
monotonic increase of particle and entropy even for the
time-symmetric recollapse of the universe. Likewise, the entropy
of a massive scalar field in the expanding and recollapsing
universe is found to be monotonically increasing using adiabatic
(WKB) solution to the classical equation of motion. Therefore, we
conclude that the entropy of the scalar field increases
monotonically throughout the evolution of the universe including
even the symmetrically recollapsing universe.

In this paper we have focused only on the scalar field in the
expanding and subsequently recollapsing universe. The other matter
of the Universe is the fermionic field. Recently, the
Liouville-von Neumann approach has been applied even to the
time-dependent fermionic field to describe correctly its
nonequilibrium evolution \cite{kim5}. Since the coupling constants
of the fermionic field depend on time as the Universe expands and
subsequently recollapses, the result of Ref. \cite{kim5} can be
employed to study the thermodynamic arrow of time due to the
fermionic field, which will be presented in a future publication.

\acknowledgements

S.P.K would like to thank Prof. R. Ruffini the warm hospitality at
the Sixth Italian-Korean Meeting on Relativistic Astrophysics.
This work was supported by the Korea Research Foundation under
contract No. 1998-001-D00364. S.P.K was also supported for the
travel expense by the Korea Science and Engineering Foundation.

\end{document}